\documentclass[aps,prl,twocolumn,preprintnumbers,superscriptaddress,amsmath,amssymb]{revtex4-1}
\usepackage{graphicx}
\usepackage{subfigure}
\usepackage{epsfig}
\usepackage{dcolumn}
\usepackage{bm}
\usepackage{ulem}
\usepackage{color}

\def\avg#1{\left\langle#1\right\rangle}

\def\be{\begin{equation}}       \def\ee{\end{equation}}
\def\bea{\begin{eqnarray}}      \def\eea{\end{eqnarray}}
\def\ba{\begin{array} }
\def\ea{\end{array} }
\def\bnum{\begin{enumerate} }
\def\enum{\end{enumerate}}

\def\nn{\nonumber}
\def\pa{\partial}
\def\=>{\Rightarrow}
\def\>{\rightarrow}
\def\A{\uparrow}
\def\V{\downarrow}

\def\eye2{Fathbb{I}}

\def\Eq#1{Eq.~(\ref{#1})}
\def\Fig#1{Fig.~\ref{#1}}

\def\Tr{\mathrm{Tr}}

\renewcommand{\>}{\rangle}

\begin{document}

\title{Topological Odd-Parity Superconductivity at Type-II 2D Van Hove Singularities}

\author{Hong Yao}
\email{yaohong@tsinghua.edu.cn}
\affiliation{Institute for Advanced Study, Tsinghua University, Beijing, 100084, China}
\affiliation{Collaborative Innovation Center of Quantum Matter, Beijing 100084, China}
\author{Fan Yang}
\affiliation{School of Physics, Beijing Institute of Technology, Beijing, 100081, China}

\begin{abstract}
We study unconventional superconductivity induced by weak repulsive interactions in 2D electronic systems at Van Hove singularity (VHS) where density of states is logarithmically divergent. We define two types of VHS. For systems at type-I VHS, weak repulsive interactions generically induce unconventional singlet pairing. However and more interestingly, for type-II VHS renormalization group (RG) analysis shows that weak repulsive interactions favor triplet pairing (e.g. $p$-wave) when the Fermi surface is not sufficiently nested. For type-II VHS systems respecting tetragonal symmetry, topological superconductivity (either chiral $p+ip$ pairing or time-reversal invariant $\mathbb{Z}_2$ $p+ip$ pairing) occurs generally. We shall also discuss relevance of our study to materials including recently discovered superconductors LaO$_{1-x}$F$_x$BiS$_2$ which can be tuned to type-II VHS by doping.
\end{abstract}
\date{\today}

\maketitle
{\bf Introduction}: Topological states of quantum matter have been among central attentions in condensed matter physics for many decades, especially after the discoveries of quantum Hall effects \cite{wen-book} and the theoretical proposal of resonating valence bond (RVB) for high temperature superconductivity (SC)\cite{anderson-87, kivelson-87, lee-nagaosa-wen-06}. More recently, a new type of time-reversal-invariant band insulators, the so-called topological insulators, have generated enormous excitements in both theoretical and experimental studies\cite{hasan-kane-10, qi-zhang-11}. It was soon realized that there exists a ``periodic table'' of noninteracting topological insulators and topological superconductors\cite{kitaev-09, schnyder-08, qi-hughes-zhang-08}.

Since the discoveries of hosting materials for 2D and 3D $\mathbb{Z}_2$ topological insulators, topological superconductivity (TSC) has attracted increasing attentions. Especially, chiral $p+ip$ topological superconductors in 2D is intriguing, partly because magnetic vortex cores in them could support Majorana zero modes\cite{volovik-99} which carry non-Abelian statistics\cite{read-green-00, ivanov-01} and may be potentially employed in realizing topological quantum computation\cite{kitaev-03, nayak-08}. Experimentally, Sr$_2$RuO$_4$ is believed by many to be chiral $p+ip$ superconductors \cite{mackenzie-03} while its experimental situations are not definitive yet. Unambiguously establishing an intrinsic topological superconductor in nature as a new state of matter is of great importance. (Certain TSC may be induced extrinsically by proximity to topologically-trivial superconductors\cite{fu-kane-08, lutchyn-10, oreg-10}).

\begin{figure}[b]
\includegraphics[scale=0.40]{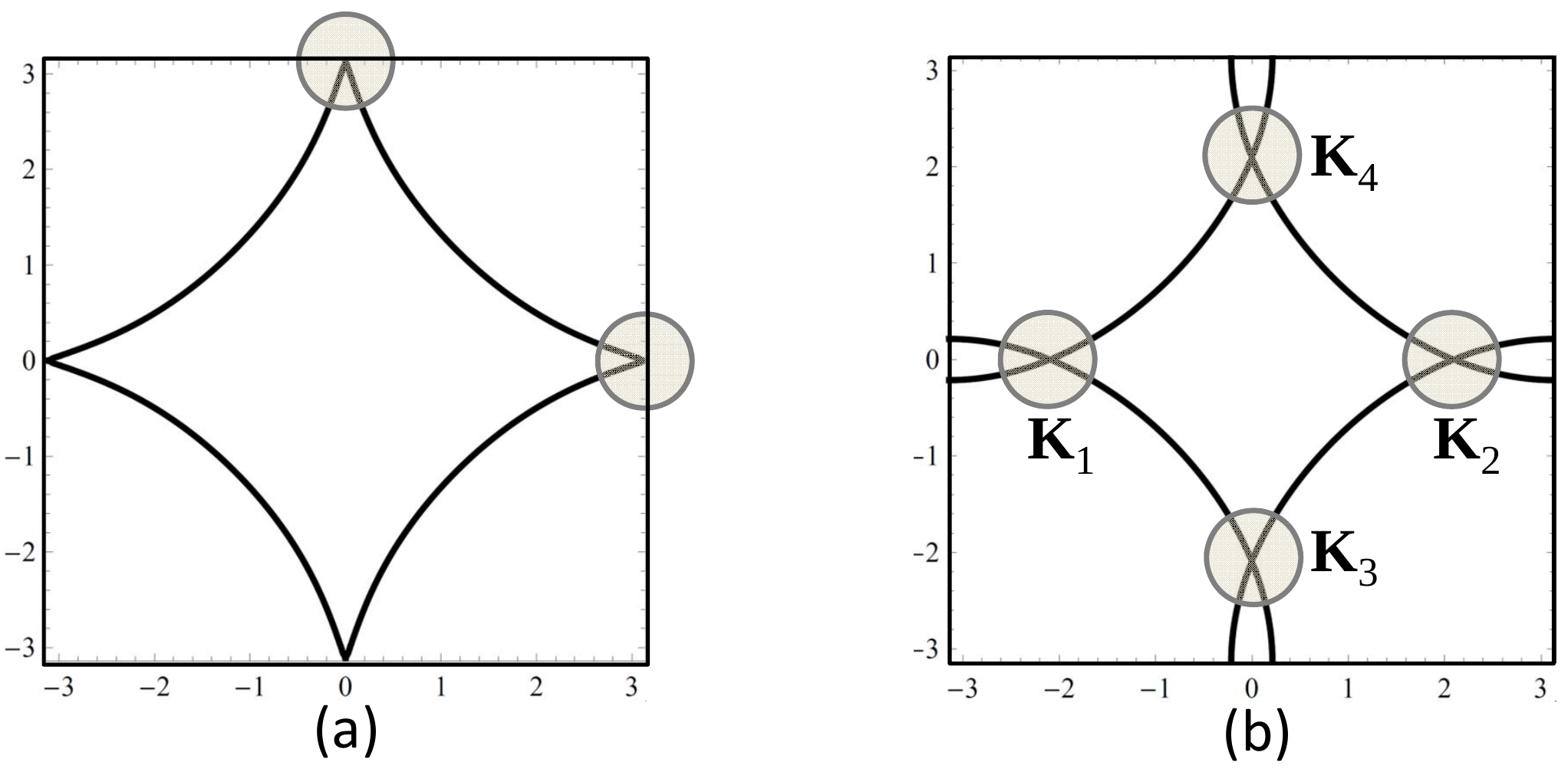}
\caption{(a) Type-I VH saddle points at $(0,\pi)$ and $(\pi,0)$ for $t_2/t_1=-0.2$ and $t_3/t_1=0.1$, as similarly realized in cuprates. (b) Type-II VH saddle points at ${\bf K}_{1,2}=(\pm K,0)$ and ${\bf K}_{3,4}=(0,\pm K)$; $K=\frac{2\pi}3$ for $t_2/t_1=-0.4$ and $t_3/t_1=0.1$.}
\label{fig1}
\end{figure}

To realize intrinsic TSC, unconventional pairing, namely sign-changing pairing among different parts of the Fermi surface (FS), is generically needed. It is widely believed that only electron-phonon coupling may not be sufficient to induce unconventional SC and repulsive interactions between electrons are often needed to form unconventional pairing. For systems with generic Fermi surfaces, it has been shown that weak repulsive interactions can induce unconventional pairing \cite{kohn-luttinger-65, zanchi-schulz-96, raghu-kivelson-scalapino-10, raghu-kivelson-11}; at the same time pairing in such systems is generically weak. However, the transition temperature to unconventional SC can be dramatically enhanced in a 2D system with FS at VHS where density of states (DOS) is logarithmically divergent\cite{schulz-87, dzyaloshinskii-87, lederer-87, furukawa-98,honerkamp-01,chubukov-12,dhlee-12,ronny-12,dhlee-13, Daniel, Hanke}. Interestingly, unconventional singlet pairing (e.g. $d$-wave) was often found in previous studies of 2D VH systems with no spin-orbit coupling \cite{schulz-87, dzyaloshinskii-87, lederer-87, furukawa-98, chubukov-12,dhlee-12,ronny-12}, which could be understood heuristically as follows. The VHS studied previously in Ref. \cite{schulz-87, dzyaloshinskii-87, lederer-87, furukawa-98, chubukov-12,dhlee-12,ronny-12} has a common feature: VH saddle points are at momenta ${\bf K}$ with ${\bf K}=-{\bf K}$, modulo reciprocal lattice vectors (such ${\bf K}$ is time-reversal-invariant). We call such saddle points as type-I VHS. Because triplet pairing $\Delta({\bf k})$ is an odd function of ${\bf k}$ [namely $\Delta(-{\bf k})=-\Delta({\bf k})$], $\Delta({\bf k})$ must vanish at type-I saddle points. Since DOS is dominantly from fermions around VH saddle points, it is expected that triplet pairing is generically suppressed in systems at type-I VHS. Note that in systems finitely away from their type-I VHS triplet pairing may be favored in the limit of weak interactions\cite{raghu-kivelson-scalapino-10}.

To realize unconventional/topological triplet pairing in systems at VHS, we study a different type of VH systems whose saddle point momenta ${\bf K}$ satisfy ${\bf K}\neq -{\bf K}$, which we call type-II VHS. The suppression of triplet pairing encountered at type-I VHS is absent for type-II VHS. Consequently, triplet pairing in general competes with singlet pairing in systems at type-II VHS. Moreover, since unconventional pairing with lowest nonzero angular momentum is $p$-wave that is a triplet pairing, we expect that triplet $p$-wave pairing can be the leading superconducting instability for weak repulsive interactions. Indeed, by RG treatment, we show that the triplet $p$-wave pairing is generally favored in systems at type-II VHS with FS not sufficiently nested (which we shall quantify below). For systems with tetragonal lattice symmetries where $p_x$ and $p_y$ pairings are degenerate forming a 2D irreducible representation of the point group, we show that topologically nontrivial $p_x+ip_y$ pairings (either chiral $p_x + ip_y$ pairing or time-reversal-invariant $\mathbb{Z}_2$ $p_x + ip_y$ pairing) gain more condensation energy than either $p_x$ or $p_y$ pairing below the transition temperature. Consequently, tetragonal systems at or close to type-II VHS may provide a promising place to look for TSC with $p_x+ip_y$ pairing.

\begin{figure}[b]
\includegraphics[scale=0.33]{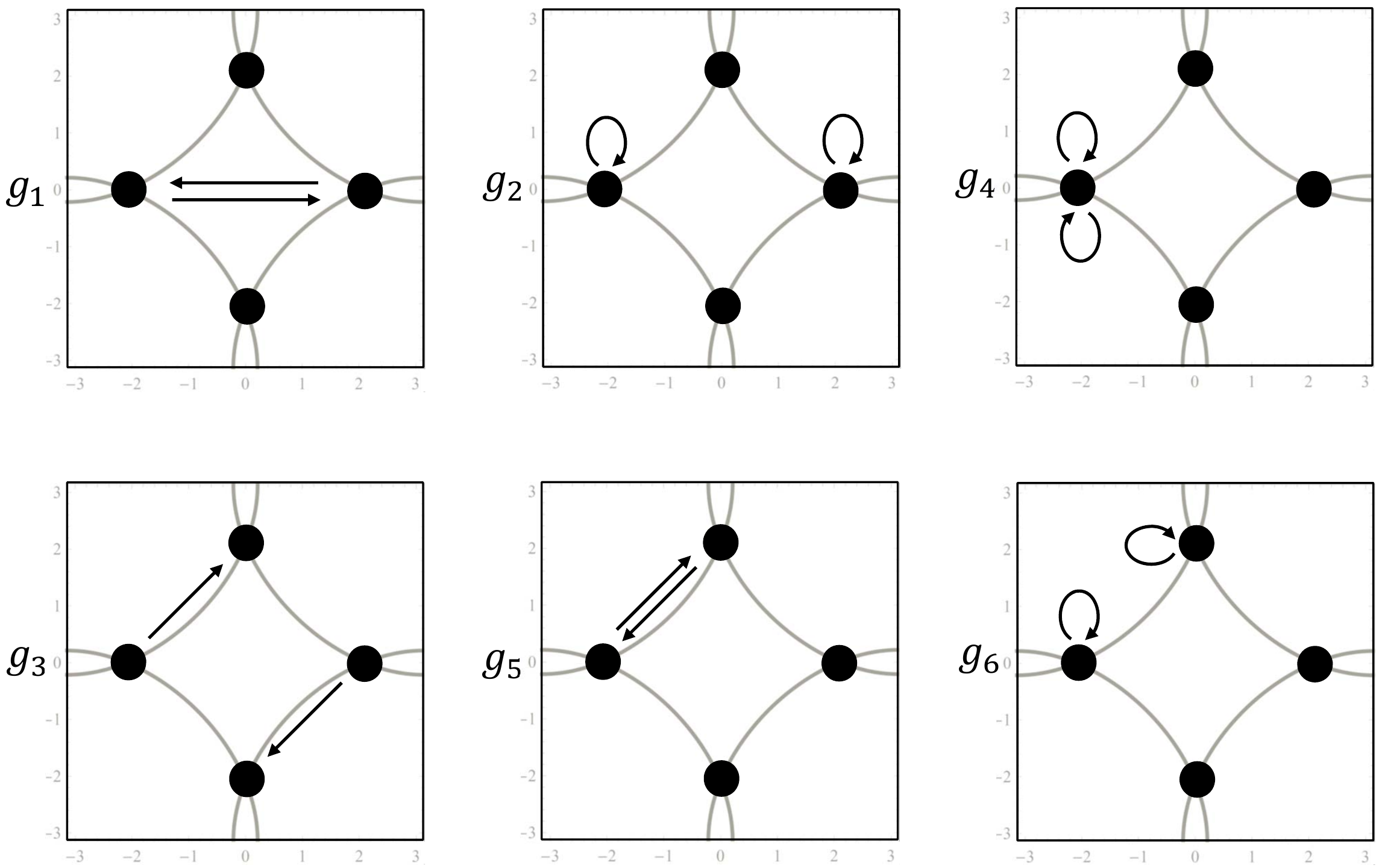}
\caption{Graphic representations of the six interactions $g_i$, $i=1,\cdots,6$. }
\label{fig2}
\end{figure}

{\bf Model and effective theory}: For simplicity, we consider the Hubbard model on the square lattice:
\bea
H=\sum_{ij\sigma}-t_{ij} c^\dag_{i\sigma} c_{j\sigma} -\sum_{i\sigma} \mu c^\dag_{i\sigma}c_{i\sigma}+\sum_i U n_{i\A}n_{i\V},~~
\eea
where $c^\dag_{i\sigma}$ are electron creation operators with spin polarization $\sigma=\A,\V$ at site $i$ and $t_{ij}=t_1$, $t_2$, $t_3$ label electron hopping between first, second, and third-neighboring sites, respectively. We assume $t_1>0$, $t_2<t_1$, and $t_3>0$. Here $U$ is the usual on-site repulsive interactions (which qualitatively simulate screened Coulomb interactions between electrons). For $|t_1+2t_2|>4t_3$ and $\mu=4t_2-4t_3$, the FS possesses type-I VH saddle points at ${\bf K}=(0,\pi)$ and $(\pi,0)$, as shown in \Fig{fig1}(a). In this paper, we shall focus the low-energy physics at type-II VHS, which is realized for $|t_1+2t_2|<4t_3$ and $\mu=(t_1+2t_2)^2/(4t_3)-2t_1$. There are four inequivalent saddle points, as shown in \Fig{fig1}(b).

In the limit of weak interactions, low energy physics is dominated by fermions around the FS. In 2D, the DOS is logarithmically divergent at VHS, namely $\rho(\omega)\sim \log(E_0/\omega)$ where $E_0$ is the order of band width and $\omega$ denotes the energy away from VHS. Moreover, the divergent DOS is contributed mainly by electrons around VH saddle points. Consequently, as a good approximation, we can neglect electrons far away from VH saddle points but focus on the electrons in patches around VH saddle points\cite{schulz-87, lederer-87, furukawa-98, honerkamp-01,chubukov-12}, which was called ``patch approximation''. Within the patch approximation, the low energy effective physics is described by the follow action:
\bea\label{eff}
&&S=\int d\tau d^2 x\sum_{a=1}^4 \Big[ \psi^\dag_{a\sigma}[\pa_\tau -\epsilon_a (i\pa_x,i\pa_y) +\mu]\psi_{a\sigma}\Big]\nn\\
&&-\sum_{a=1}^4\left[\frac{g_1}2\psi^\dag_a\psi^\dag_{\bar a}\psi_a\psi_{\bar a}  +\frac{g_2}2\psi^\dag_a\psi^\dag_{\bar a}\psi_{\bar a}\psi_a\right]-\sum_{a=1}^4\frac{g_4}2 \psi^\dag_a\psi^\dag_a\psi_a\psi_a \nn\\
&&-g_3\left[(\psi^\dag_{1\A}\psi^\dag_{2\V}+\psi^\dag_{2\A}\psi^\dag_{1\V})(\psi_{3\V}\psi_{4\A} +\psi_{4\V}\psi_{3\A})+h.c.\right]\nn\\
&&-\sum_{a=1}^2\sum_{b=3}^4 \left[g_5\psi^\dag_a\psi^\dag_b\psi_a\psi_b  +g_6\psi^\dag_a\psi^\dag_b\psi_b\psi_a\right],
\eea
where $\psi_{a\sigma}$ is annihilation operators of electrons with patch indices $a=1,\cdots,4$ and spin polarization $\sigma=\A,\V$. Here $\bar a$ labels patches with momenta opposite to $a$. Short-range interactions $g_i$ are represented graphically in \Fig{fig2}. A spin sum with the indices $\sigma,\sigma',\sigma',\sigma$ is implicit in the above expressions for interactions $g_1,g_2,g_4,g_5,g_6$. Note that $g_3$ scattering occurs only in the singlet channel because of the tetragonal lattice symmetries\cite{footnote2}. Here, $\epsilon_a(k_x,k_y)$ is the dispersion of VH electrons expanded up to quadratic terms around saddle point $a$:  $\epsilon_{1,2}({\bf k})\approx -\frac{k_x^2}{2m_x}+\frac{k_y^2}{2m_y}$ and $\epsilon_{3,4}({\bf k})\approx \frac{k_x^2}{2m_y}-\frac{k_y^2}{2m_x}$. From the lattice model with $t_1$, $t_2$ and $t_3$, we obtain  $m_x=2t_3/[(4t_3-2t_2-t_1)(4t_3+2t_2+t_1)]$ and $m_y=2t_3/[(4t_3-2t_2)(4t_3+2t_2+t_1)]$. Chemical potential $\mu=0$ in \Eq{eff} represents systems exactly at VHS. We expect that short-range interactions in the effective theory can qualitatively capture the low energy physics of realistic materials at VHS because long-range interactions may be sufficiently screened by electrons at Fermi level with divergent DOS.

To determine which kind of FS instability occurs as temperature decreases, we study how interactions flow using one-loop RG equations derived from gradually integrating out electrons at high-energy. In doing so, it is essential to know susceptibilities of various particle-hole and particle-particle channels at low-energy $\omega$ in the noninteracting limit; especially,  susceptibilities in the Cooper channel have log-square behavior.
\bea
\chi^\textrm{pp}_{\bf 0}(\omega) &\approx&\lambda\log^2\frac{E_0}{\omega},~~\chi^\textrm{ph}_{\bf 0}(\omega) \approx 2\lambda\log\frac{E_0}{\omega},\\
\chi^\textrm{pp}_{{\bf Q}_1}(\omega) &\approx&\lambda \gamma \log^2\frac{E_0}{\omega},~~\chi^\textrm{ph}_{{\bf Q}_1}(\omega) \approx2\lambda\gamma\log\frac{E_0}{\omega},\\
\chi^\textrm{pp}_{{\bf Q}_2}(\omega) &\approx& 2\lambda\gamma_1\log\frac{E_0}{\omega},~~\chi^\textrm{ph}_{{\bf Q}_2}(\omega) \approx 2 \lambda \gamma_2 \log\frac{E_0}{\omega},
\eea
where $\lambda=\sqrt{m_xm_y}/(4\pi^2)$ is the parameter characterizing the DOS per patch $\rho(\omega)\approx 2\lambda \log(E_0/\omega)$, and ${\bf Q}_1=2{\bf K}_1$ and ${\bf Q}_2={\bf K}_1-{\bf K}_3$ are two inequivalent momenta connecting different saddle points. $\gamma_1\approx \frac{1+\kappa}{2\sqrt{\kappa}}$ and $\gamma_2\approx \frac{2\sqrt{\kappa}}{1+\kappa} \log|\frac{\kappa+1}{\kappa-1}|$, where $\kappa=\frac{m_y}{m_x}$ is the ratio of masses of the VH saddle points. The mass ratio $\kappa$ characterizes how perfect different patches of the FS are nested by ${\bf Q}_2$; $\kappa = 1$ or $m_x=m_y$ represents perfect nesting. $\gamma$ is a constant with $0<\gamma<1$. In the following, we shall show by RG analysis that the pairing symmetry of SC induced by weak repulsive interactions mainly depends on the mass ratio $\kappa$ while its dependence on $\gamma$ is negligible. When $\kappa>\kappa_c$, namely the FS is not sufficiently nested, the most favored pairing symmetry of SC is in the triplet $p$-wave channel.

{\bf RG equations and triplet pairing}: RG flow equations can be obtained by extending approaches developed in previous studies\cite{schulz-87, lederer-87, furukawa-98, chubukov-12}; we use a Wilson RG flow parameterized by a decreasing energy cutoff $E_0$. Since all $g_i$ are marginal at the tree level, we go to one-loop RG, which is expected to capture the leading behavior at low energies when the couplings are weak. With logarithmic accuracy, using $y\equiv \log^2(E_0/\omega)\sim \chi^\textrm{pp}_{\bf 0}$ as the RG flow parameter, we obtain the following RG equations:
\bea
\dot g_1&=&-2g_1g_2-2g^2_3+2d_2 g_1(g_2-g_1)+2d_1 (g_1g_4+g_5^2), ~~ \\
\dot g_2&=&-(g_1^2+g_2^2)-2g^2_3+d_2g_2^2 \nn\\
 &&~~~+2d_1[g_4(g_1-g_2)+2g_6(g_5-g_6)], \\
\dot g_3&=& g_3[-2(g_1+g_2)+2d_3(2g_6-g_5)],\\
\dot g_4&=&-d_4g_4^2 +d_1[g_1^2+2g_2(g_1-g_2)+g_4^2\nn\\
&&~~~+2g_5^2+4g_6(g_5-g_6)],\\
\dot g_5&=&g_5[-2d_5g_6+2d_3(g_6-g_5)+2d_1(g_1+g_4)],\\
\dot g_6&=&-d_5(g^2_5+g^2_6) +d_3[g^2_6+g^2_3] \nn\\
&&~~~ +2d_1[(g_2+g_4)(g_5-g_6)+g_6(g_1-g_2)],
\eea
where dimensionless couplings are used by introducing $g_i\to \lambda g_i$ and $\dot g_i=\frac{dg_i}{dy}$. Here $d_1(y)=\pa\chi^\textrm{ph}_{{\bf 0}}/\pa\chi^\textrm{pp}_{{\bf 0}}$, $d_2(y)=\pa\chi^\textrm{ph}_{{\bf Q}_1} /\pa\chi^\textrm{pp}_{{\bf 0}}$, $d_3(y)=\pa \chi^\textrm{ph}_{{\bf Q}_2}/\pa\chi^\textrm{pp}_{{\bf 0}}$, $d_4(y)=\pa\chi^\textrm{pp}_{{\bf Q}_1}/\pa\chi^\textrm{pp}_{{\bf 0}}$, $d_5(y)=\pa \chi^\textrm{pp}_{{\bf Q}_2}/\pa\chi^\textrm{pp}_{{\bf 0}}$. It is clear that $d_i(y)$ are decreasing functions of $y$ for $y\ge 0$ and they have asymptotic behaviors: $d_i(y)\to 1$ as $y\to 0$, $i=1,\cdots,5$; $d_1(y)\to \frac{1}{\sqrt{y}}$, $d_2(y)\to \frac{\gamma}{\sqrt{y}} $, $d_3(y)\to \frac{\gamma_2}{\sqrt{y}} $, $d_4(y)\to \gamma $, and $d_5(y) \to \frac{\gamma_1}{\sqrt{y}}$ as $y\to \infty$. To qualitatively capture the low energy physics, we model $d_i(y)$ as follows: $d_1(y)\approx \frac{1}{\sqrt{1+y}}$, $d_2(y)\approx \frac{\gamma}{\sqrt{\gamma^2+y}}$, $d_3(y)\approx \frac{\gamma_2}{\sqrt{\gamma^2_2+y}}$, $d_4(y)=\frac{1+\gamma y}{1+y}$, and $d_5(y)\approx \frac{\gamma_1}{\sqrt{\gamma_1^2+y}}$. From the RG equations, $g_i$ flow to strong coupling limit as $y$ approaches the instability threshold $y_c$. Close to $y_c$, $g_i(y)\approx \frac{G_i}{y_c-y}$, where $G_i$ are constants that depend on $\gamma$, $\kappa$, and the initial interactions $g_i(0)$. For the Hubbard model which we consider, $g_i(0)=\lambda U$.

\begin{figure}[b]
\includegraphics[scale=0.55]{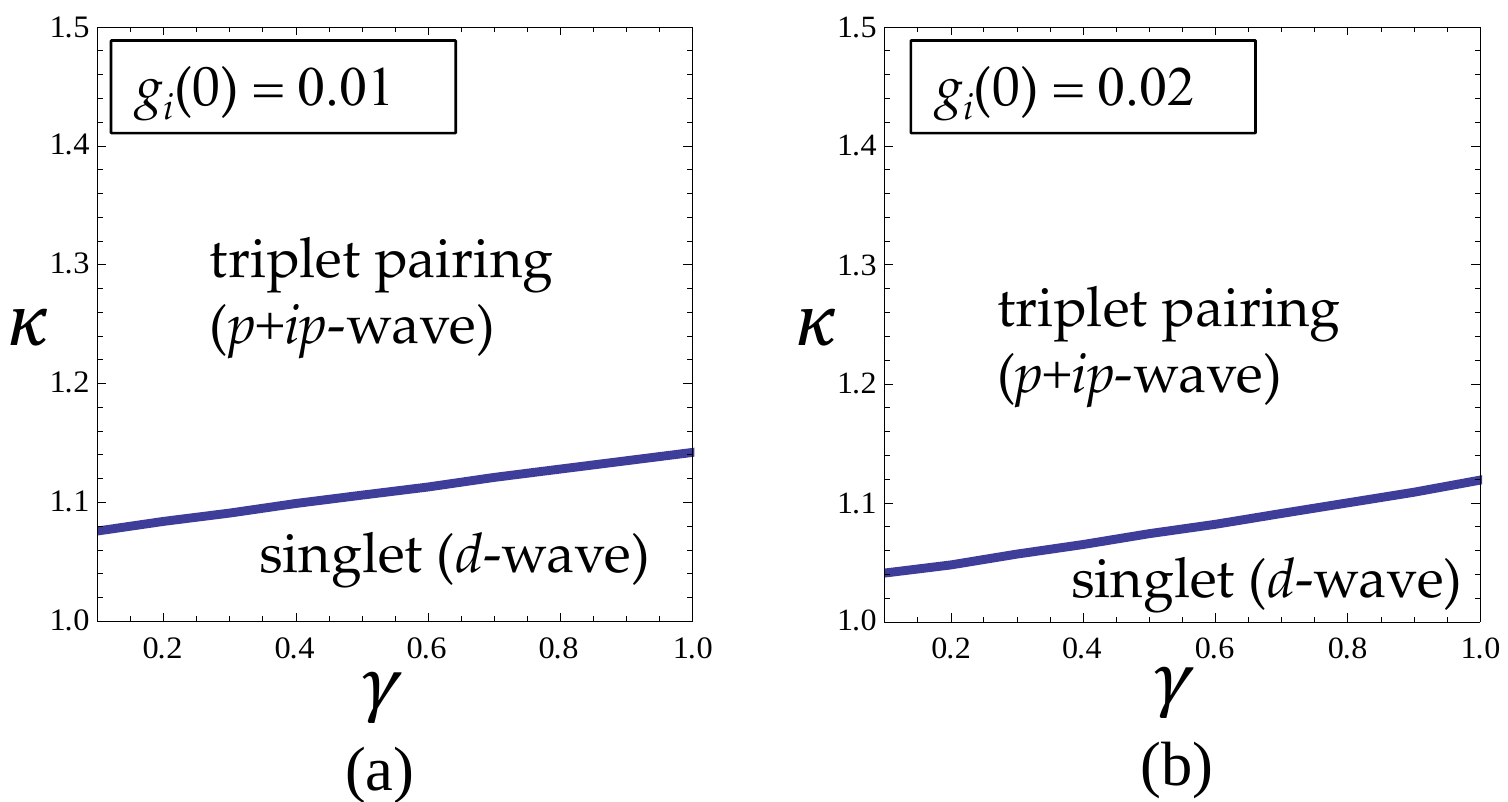}
\caption{ The phase diagrams as a function of $\gamma$ and $\kappa$ for $g_i(0)=0.01$ and $0.02$ (or $U/t_1\sim 0.01$ and $0.02$), respectively.}
\label{fig3}
\end{figure}

Normally, knowing one-loop RG equations does not directly solve for the strong coupling fixed point. Nonetheless, the qualitative phase diagram may be obtained from computing various susceptibilities, which diverge close to $y_c$ as $(y_c-y)^\alpha$. The actually broken symmetry occurring at the phase transition has the most negative $\alpha$. We obtain $\alpha$ of leading possible broken symmetries:  $\alpha_{p\textrm{-SC}}=2(G_2-G_1)$, $\alpha_{s\textrm{-SC}}=2(G_2+G_1+2G_3)$, $\alpha_{d\textrm{-SC}}=2(G_2+G_1-2G_3)$, $\alpha_{{\bf Q}_1\textrm{-FFLO}}=2G_4 d_5(y_c)$, $\alpha_{{\bf Q}_2\textrm{-SDW}}=2(G_3-G_6)d_3(y_c)$, and $\alpha_\textrm{FM}=-2(G_1+G_4+G_5)d_1(y_c)$, where the subscripts represent $p$-wave triplet pairing, $s$-wave singlet pairing, $d$-wave singlet pairing, the FFLO pairing at finite momentum ${\bf Q}_1$, spin density wave (SDW) at ${\bf Q}_2$, and ferromagnetic state, respectively. In the weak interaction limit and without perfect FS nesting, $\alpha$ in particle-hole channels are not as negative as the ones in particle-particle channels; consequently, our discussions below will focus on pairing instability even though we compute susceptibilities in both pairing and particle-hole channels \cite{footnote4}.

\begin{figure}
\includegraphics[scale=0.4]{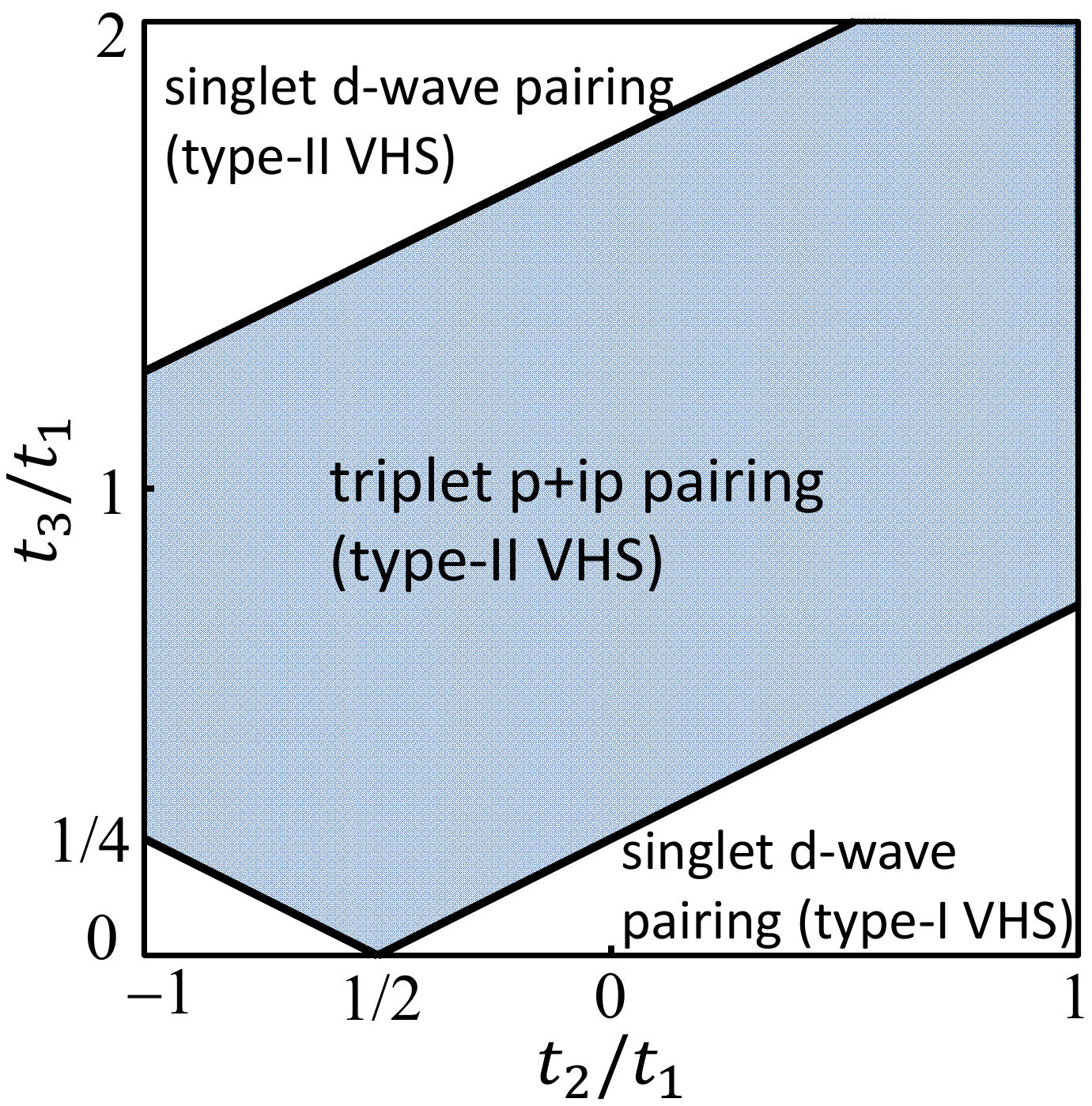}
\caption{The schematic phase diagram of the Hubbard model at VHS and with weak repulsive interactions $U$. For type-I VHS, the pairing is always in singlet channel and has $d$-wave symmetry. While for type-II VHS, topological triplet pairing is favored in a major part of the phase diagram. }
\label{fig4}
\end{figure}

We solve the RG flow equations [Eq. (6-11)] numerically by discretizing the differential equations to find out the flow of $g_i$ with the parameter $y$ and to identify the most relevant interaction channel, from which we obtain the phase diagram as a function of $\gamma$ and $\kappa$ for different initial interactions. For $g_i(0)=0.01$ and $0.02$, the phases diagrams are shown in \Fig{fig3} (a) and (b), respectively. For $\kappa>\kappa_c$, the triplet $p$-wave pairing wins over the singlet $d$-wave pairing. We note that the critical value $\kappa_c$ depends on $\gamma$ and also bare interactions $g_i(0)$. With increasing interactions $g_i(0)$, the region of triplet pairing is enlarged. 
Moreover, triplet $p$-wave pairing is favored unless $\kappa$ is extremely close to 1 ($\kappa=1$ represents the perfect nesting limit). 
This is one of the central results of this present work. 
Converting $\gamma$ and $\kappa$ back to hopping parameters $t_1$, $t_2$ and $t_3$ of the Hubbard model, we obtain a schematic phase diagram as shown in \Fig{fig4}. It is clear that triplet $p$-wave pairing is the dominant instability in a major part of parameter space realizing type-II VHS, and without good Fermi surface nesting. Note that perfect FS nesting only occurs at the isolated point $(t_2=0, t_3=0)$ in the phase diagram, which is actually a type-I VHS.

It is worth to point out that in the limit that $g_3=g_5=g_6=0$, namely patches ${\bf K}_1$ and ${\bf K}_2$ are not interacting with  patches ${\bf K}_3$ and ${\bf K}_4$, the pairing symmetry is always in the triplet channel for arbitrary $\gamma$, $\kappa$, and $g_i(0)$. This can be understood through the RG flow of $g_1$: $\dot g_1=g_1[-2g_2+2d_1 (g_2-g_1)+2d_2 g_4]$;
it is clear that $g_1>0$ if it starts out positive. Consequently, $G_1>0$ and $\alpha_{p\textrm{-SC}}<\alpha_{s\textrm{-SC}}$, namely the susceptibility of triplet pairing diverges faster than singlet for arbitrary $\gamma$, $\kappa$, and $g_i(0)>0$.

{\bf Topological superconductivity}: We have shown that triplet $p$-wave pairing is favored for systems at type-II VHS in the limit of weak repulsive interactions. To fully characterize a triplet pairing, one need to specify ${\bf d}_{\bf k}$, defined through $\avg{\psi^\dag_{\bf k} \psi^\dag_{-{\bf k}}}\propto i\sigma^y {\bf d}_{\bf k} \cdot {\boldsymbol\sigma} $.  On one hand, for a system without spin-orbit coupling, the free energy is independent on the global direction of ${\bf d}_{\bf k}$. In 2D, the spin-rotational symmetry can be spontaneously broken at zero temperature, which select a global direction of ${\bf d}_{\bf k}$. On the other hand, due to the crystalline symmetry, $p_x$ and $p_y$ pairings are degenerate and have identical pairing susceptibility. Either $p_x$ or $p_y$ pairing cannot fully gap out the FS while $p_x+ip_y$ pairing can, we expect that the $p_x+ip_y$ pairing gain more condensation energy than either $p_x$ or $p_y$ pairing below the transition temperature\cite{ksun-10}. Indeed, from the analysis of Ginzburg-Landau free energy (see Appendix), it is shown that either of the following two kinds of $p_x+ip_y$ pairings minimizes the free energy:  (i) ${\bf d}_{\bf k} \propto (k_x+ik_y)\hat {\bf e}$, for which the pairing is chiral and breaks time reversal symmetry; consequently, the system supports robust gapless Majorana edge states on its boundary and a non-Abelian Majorana zero mode in the core of a magnetic $hc/4e$ half-vortex\cite{ivanov-01}. (ii) ${\bf d}_{\bf k}\propto k_x\hat {\bf e}_1+k_y\hat {\bf e}_2$ with $\hat {\bf e}_1\perp \hat {\bf e}_2$, for which the pairing between two spin-``up'' electrons is $p_x+ip_y$ and between two spin-``down'' electrons is $p_x-ip_y$ (spin quantization axis is ${\bf e}_1\times {\bf e}_2$); such pairing preserves time reversal symmetry and is topologically nontrivial with $Z_2$ topological invariant \cite{schnyder-08,kitaev-09,qi-raghu-hughes-zhang-09}; the system carries helical Majorana modes along its edge. These two topological pairings have degenerate energy for a system without spin-orbit coupling. In real materials where spin-orbit coupling is always present, one of the two topological pairings will be favored below the superconducting phase transition.

{\bf Discussions and concluding remarks}: A new family of layered superconductors LaO$_{1-x}$F$_x$BiS$_2$ were discovered recently\cite{experiment1,experiment2}. According to {\it ab initio} calculations\cite{kuroki-12}, the relevant orbitals close to the Fermi level are $p_x$ and $p_y$ of Bi atoms which form a square lattice in each layer. By tuning $x$, the FS goes through a type-II VHS where the FS has no nesting\cite{kuroki-12,yang-13}. The RG analysis in the present work indicates that the dominant pairing is in the triplet channel while assuming that its SC is mainly induced by weak repulsive interactions. Moreover, employing a two-orbital Hubbard model which is expected to capture the main physics of this material, RPA-type analysis around the type-II VHS shows that weak short-range repulsive interactions induced SC is in the triplet channel and the tendency towards triplet pairing can be significantly enhanced by spin-orbit coupling\cite{yang-13}. Recently reported $H_{c_2}$ measurements of this material\cite{haihu} provides preliminary evidences for triplet pairings  even though its pairing symmetry remains largely unknown.

Besides tetragonal systems, type-II VHS in 2D may also occur in systems with  hexagonal symmetry, for which there are generally six type-II VH saddle points. Band-structure calculations show that such type-II VHS can be realized in the doped BC$_3$\cite{xichen-13}. For such type-II VHS, RG flow equations are more complex and studies of its phase diagram will be presented in the future\cite{xichen-13}. In the end, we would like to mention that in real materials, the properties of VHS can be impacted by disorder\cite{Jones}. Whether such impact will bring any effect on the pairing symmetry is currently not obvious. We leave this topic for future study.

{\it Acknowledgement}: We would like to thank Steve Kivelson and Shao-Kai Jian for helpful discussions. This work is supported in part by the National Thousand-Young-Talents Program and the NSFC under Grant No. 11474175 at Tsinghua University (HY), and by NSFC under Grant Nos. 11274041 and 11334012 and the NCET program under Grant No. NCET-12-0038 at BIT (FY).

\begin{widetext}
\subsection{Appendix: Ginzburg-Landau Free Energy of Triplet $p$-wave Pairings}
To investigate the interactions between two degenerate triplet pairing order parameters $\phi_1({\bf k}) \hat {\bf e}_1$ and $\phi_2({\bf k}) \hat {\bf e}_2$, where $\phi_1({\bf k})$ and $\phi_2({\bf k})$ comprise the two-dimensional irreducible representation of the point group and $\hat {\bf e}_1$ and $\hat {\bf e}_2$ are two unit vectors, we derive the Ginzburg-Landau free energy in terms of the two order parameters by the linked cluster expansion. [The simplest form of $\phi_1({\bf k})$ and $\phi_2({\bf k})$ are $k_x$ and $k_y$, respectively.] The signs of certain quartic terms in the Ginzburg-Landau free energy can tell us whether the two order parameters are repulsive or attractive. The partition function in the path integral formulism is $Z=\int \mathcal{D}\bar\psi\mathcal{D}\psi \exp(-\int d\tau \mathcal{L}[\bar\psi,\psi])$, where
\bea
\mathcal{L}=\sum_{\bf k} \psi^\dag_{\bf k} \left[\pa_\tau-\epsilon(\bf k)+\mu\right]\psi_{\bf k} -\sum_{{\bf k},{\bf k'}} \psi^\dag_{\bf k}\psi^\dag_{-{\bf k}}V({\bf k},{\bf k'})\psi_{-{\bf k'}}\psi_{\bf k'},
\eea
where $V({\bf k},{\bf k'})$ is the effective pairing interactions of electrons close to the FS after high energy degree of freedoms have been integrated out. Assume that the most favored pairing symmetry is triplet odd-parity pairing and there are two degenerate odd-parity $p$-wave pairings with normalized pairing functions $\phi_1({\bf k})$ and $\phi_2({\bf k})$, where $\int d^2 {\bf k} \phi_i^2({\bf k}) =1$ and $\int d^2 {\bf k} \phi_1({\bf k}) \phi_2({\bf k}) =0$. Taking a general pairing $\boldsymbol\Delta({\bf k})=\Delta_1\phi_1({\bf k})\hat {\bf e}_1+\Delta_2\phi_2({\bf k})\hat {\bf e}_2$, where $\Delta_1$ and $\Delta_2$ are the two pairing parameters, the partition function is given by
\bea
\mathcal{L}&=&\sum_{\bf k} (\psi^\dag_{\bf k\A},\psi^\dag_{\bf k\V},\psi_{-\bf k\A},\psi_{-\bf k\V})
\left(\ba{cc}
G^{-1}_p & i\sigma^y\boldsymbol\sigma\cdot\boldsymbol\Delta({\bf k})  \\
\boldsymbol\sigma \cdot\boldsymbol\Delta^\ast({\bf k}) (-i\sigma^y) & G^{-1}_h
\ea\right)
\left(\ba{c}
\psi_{\bf k\A}\\ \psi_{\bf k\V} \\ \psi^\dag_{-\bf k\A} \\ \psi^\dag_{-\bf k\V}
\ea\right)\nn\\
&+&\sum_{{\bf k},{\bf k'}}\left[|\Delta_1|^2\phi_1({\bf k}) V^{-1}({\bf k},{\bf k'})\phi_1({\bf k'})+|\Delta_2|^2 \phi_2({\bf k}) V^{-1}({\bf k},{\bf k'})\phi_2({\bf k'})\right],
\eea
where $G_{p/h}=\frac1{i\omega\mp [\epsilon({\bf k})-\mu]}$ are the Green's functions of electrons and holes with Matsubara frequency $\omega$. By integrating out the fermions, we obtain the action in terms of the order parameters:
\bea\label{eq16}
\tilde {\mathcal L}=\Tr\log
\left(\ba{cc}
G^{-1}_p & i\sigma^y\boldsymbol\sigma\cdot\boldsymbol\Delta({\bf k})  \\
\boldsymbol\sigma \cdot\boldsymbol\Delta^\ast({\bf k})(-i\sigma^y) & G^{-1}_h
\ea\right)
+\frac{1}{\lambda_0}\left(|\Delta_1|^2+|\Delta_2|^2\right),
\eea
where $\lambda_0$ is the largest positive eigenvalue of the interaction $V({\bf k},{\bf k'})$ whose eigenfunctions are $\phi_1({\bf k})$ and $\phi_2({\bf k})$. By expanding the $\Tr\log$ in \Eq{eq16} to the quartic order of $\Delta_i$, we obtain
\bea
\tilde {\mathcal L}&=&{\mathcal L}_0+2\Tr\left[ G_p i\sigma^y\boldsymbol\sigma\cdot\boldsymbol\Delta({\bf k}) G_h \boldsymbol\sigma \cdot\boldsymbol\Delta^\ast({\bf k})(-i\sigma^y)\right] +\frac{1}{\lambda_0}\left(|\Delta_1|^2+|\Delta_2|^2\right) \nn\\
&&~~~~+2\Tr\left[ G_p i\sigma^y\vec \sigma\cdot\vec \Delta({\bf k}) G_h \boldsymbol\sigma \cdot\boldsymbol\Delta^\ast({\bf k})(-i\sigma^y) G_p i\sigma^y\boldsymbol\sigma \cdot \boldsymbol\Delta({\bf k}) G_h \boldsymbol\sigma \cdot \boldsymbol\Delta^\ast({\bf k})(-i\sigma^y) \right],\\
&=&{\mathcal L}_0+r\left(|\Delta_1|^2+|\Delta_2|^2\right)+u_1(|\Delta_1|^2+|\Delta_2|^2)^2 + 2u_2|\Delta_1|^2|\Delta_2|^2+u_3|\Delta_1\Delta^\ast_2+\Delta^\ast_1\Delta_2|^2 +u_4|\Delta_1\Delta^\ast_2-\Delta^\ast_1\Delta_2|^2,
\eea
where
\bea
r&=&\Tr \left(G_p G_h\right)+1/\lambda_0,\\
u_1&=&\Tr \left(G_p G_h G_p G_h \phi_1^4\right)=\Tr \left(G_p G_h G_p G_h \phi_2^4\right),\\
u_2&=&\Tr \left[G_p G_h G_p G_h \phi_1^2\phi_2^2\right]-u_1,\\
u_3&=&\Tr\left[G_pG_hG_pG_h\phi_1^2\phi_2^2 (\hat {\bf e}_1\cdot\hat {\bf e}_2)^2 \right],\\
u_4&=&\Tr\left[G_pG_hG_pG_h\phi_1^2\phi_2^2 (\hat {\bf e}_1\times\hat{\bf e}_2)^2 \right].
\eea
It is clear that $u_1>0,u_2<0,u_3>0,u_4>0$ and $u_1+u_2=u_3+u_4$. The $u_3$ and $u_4$ terms represent the real and imaginary parts of $\Delta_1\Delta_2^\ast$, respectively. Since $u_2<0$, there are two ways to minimize the $u_2$, $u_3$, and $u_4$ simultaneously: (i) $\hat {\bf e}_1 \parallel \hat {\bf e}_2$ and $\Delta_2=\pm i \Delta_1$, which corresponds to chiral $p+ip$ pairing;  (ii) $\hat {\bf e}_1\perp \hat {\bf e}_2$ and $\Delta_2=\pm \Delta_1$, which represents time reversal invariant $p+ip$ pairing. These two kinds of $p+ip$ pairings have the same free energy and  are degenerate. But the presence of spin-orbit couplings in real materials can lift the degeneracy and select the one with lower free energy.
\end{widetext}


\begin{thebibliography}{33}

\bibitem{wen-book} X.-G. Wen, {\it Quantum Field Theory of Many-body Sys
tems}, (Oxford University Press, New York, 2004).

\bibitem{anderson-87} P. W. Anderson, Science {\bf 235}, 1196 (1987).

\bibitem{kivelson-87}S. A. Kivelson, D. S. Rokhsar, and J. P. Sethna, Phys. Rev. B {\bf 35}, 8865 (1987).

\bibitem{lee-nagaosa-wen-06}P. A. Lee, N. Nagaosa, and X.-G. Wen, Rev. Mod. Phys. {\bf 78}, 17 (2006).

\bibitem{hasan-kane-10} M. Z. Hasan and C. L. Kane, Rev. Mod. Phys. {\bf 82}, 3045 (2010).

\bibitem{qi-zhang-11} X.-L. Qi and S.-C. Zhang, Rev. Mod. Phys. {\bf 83}, 1057 (2011).

\bibitem{kitaev-09} A. Kitaev, AIP Conf. Proc. {\bf 1134}, 22 (2009).

\bibitem{schnyder-08} A. P. Schnyder, S. Ryu, A. Furusaki, and A. W. W. Ludwig, Phys. Rev. B {\bf 78}, 195125 (2008).

\bibitem{qi-hughes-zhang-08} X.-L. Qi, T. L. Hughes, and S.-C. Zhang, Phys. Rev. B {\bf 78}, 195424 (2008).

\bibitem{volovik-99} G. E. Volovik, JETP Lett. {\bf 70}, 609 (1999).

\bibitem{ivanov-01} D. A. Ivanov, Phys. Rev. Lett. {\bf 86}, 268 (2001).

\bibitem{read-green-00} N. Read and D. Green, Phys. Rev. B {\bf 61}, 10267 (2000).

\bibitem{kitaev-03} A. Y. Kitaev, Ann. Phys. {\bf 303}, 2 (2003).

\bibitem{nayak-08} C. Nayak, S. H. Simon, A. Stern, M. Freedman, and S. D. Sarma, Rev. Mod. Phys. {\bf 80}, 1083 (2008).

\bibitem{mackenzie-03} A. P. Mackenzie, and Y. Maeno, Rev. Mod. Phys. {\bf 75}, 657 (2003).

\bibitem{fu-kane-08} L. Fu and C. L. Kane, Phys. Rev. Lett. {\bf 100}, 096407 (2008).

\bibitem{lutchyn-10} R. M. Lutchyn, J. D. Sau, and S. Das Sarma, Phys. Rev. Lett.
{\bf 105}, 077001 (2010).

\bibitem{oreg-10} Y. Oreg, G. Refael, and F. von Oppen, Phys. Rev. Lett. {\bf 105}, 177002 (2010).

\bibitem{kohn-luttinger-65} W. Kohn and J. M. Luttinger, Phys. Rev. Lett. {\bf 15}, 524 (1965).

\bibitem{zanchi-schulz-96} D. Zanchi and H. J. Schulz, Phys. Rev. B {\bf 54}, 9509 (1996).

\bibitem{raghu-kivelson-scalapino-10} S. Raghu, S. A. Kivelson, and D. J. Scalapino, Phys. Rev. B {\bf 81}, 224505 (2010).

\bibitem{raghu-kivelson-11} S. Raghu and S. A. Kivelson, Phys. Rev. B {\bf 83}, 094518 (2011).

\bibitem{dzyaloshinskii-87} I. E. Dzyaloshinskii, JETP Lett. {\bf 46}, 118 (1987).

\bibitem{schulz-87} H. J. Schulz, Europhys. Lett. {\bf 4}, 609 (1987).

\bibitem{lederer-87} P. Lederer, G. Montambaux, and D. Poilblanc, J. Phys. (Paris)
{\bf 48}, 1613 (1987).

\bibitem{furukawa-98}  N. Furukawa, T. M. Rice, and M. Salmhofer, Phys. Rev. Lett. {\bf 81}, 3195 (1998).

\bibitem{honerkamp-01} C. Honerkamp and M. Salmhofer, Phys. Rev. Lett. {\bf 87}, 187004 (2001).

\bibitem{chubukov-12} R. Nandkishore, L. S. Levitov, and A. V. Chubukov, Nat. Phys. {\bf 8}, 158 (2012).

\bibitem{dhlee-12} W.-S. Wang, Y.-Y. Xiang, Q.-H. Wang, F. Wang, F. Yang, and D.-H. Lee, Phys. Rev. B {\bf 85}, 035414 (2012).

\bibitem{ronny-12} M. L. Kiesel, C. Platt, W. Hanke, D. A. Abanin, and R. Thomale, Phys. Rev. B {\bf 86}, 020507(R) (2012).

\bibitem{dhlee-13} Y.-Y. Xiang, W.-S. Wang, Q.-H. Wang, and D.-H. Lee, Phys. Rev. B {\bf 86}, 024523 (2012).

\bibitem{Daniel} D. D. Scherer, M. M. Scherer, G. Khaliullin, C. Honerkamp, and B. Rosenow, Phys. Rev. B {\bf 90}, 045135 (2014).

\bibitem{Hanke} T. Hanke, S. Sykora, R. Schlegel, D. Baumann, L. Harnagea, S. Wurmehl, M. Daghofer, B. B¨¹chner, J. van den Brink, and C. Hess, Phys. Rev. Lett. {\bf 108}, 127001(2012).

\bibitem{footnote2} For instance, reflection symmetry of $x\to -x$ would exchange patch ${\bf K}_1$ and ${\bf K}_2$, which requires that the interaction in the triplet channel vanishes.

\bibitem{ksun-10} M. Cheng, K. Sun, Victor Galitski, and S. Das Sarma, Phys. Rev. B {\bf 81}, 024504 (2010).

\bibitem{footnote4} We defer the cases of perfectly nested Fermi surface with type-II VHS to a future work.


\bibitem{qi-raghu-hughes-zhang-09} X.-L. Qi, T. L. Hughes, S. Raghu, and S.-C. Zhang, Phys. Rev. Lett. {\bf 102}, 187001 (2009).

\bibitem{experiment1} Y. Mizuguchi, H. Fujihisa, Y. Gotoh, K. Suzuki, H. Usui, K. Kuroki, S. Demura, Y. Takano, H. Izawa, and O. Miura, 
    Phys. Rev. B {\bf 86}, 220510 (2012).

\bibitem{experiment2} Y. Mizuguchi, S. Demura, K. Deguchi, Y. Takano, H. Fujihisa, Y. Gotoh, H. Izawa, and O. Miura, 
    J. Phys. Soc. Japan {\bf 81}, 114725 (2012).

\bibitem{kuroki-12} H. Usui, K. Suzuki, and K. Kuroki, Phys. Rev. B {\bf 86}, 220501(R)(2012).

\bibitem{yang-13} Yi-Fan Jiang, Fan Yang, and Hong Yao, to be published (2014).

\bibitem{haihu}  S. Li, H. Yang, D.-L. Fang, Z.-Y Wang, J. Tao, X. Ding,
and H.-H. Wen, Sci. China-Phys Mech Astron {\bf 56}, 2019 (2013).

\bibitem{xichen-13} X. Chen, Y. Yao, H. Yao, F. Yang, and J. Ni, arXiv:1404.3346.

\bibitem{Jones} See for example, W. Jones and N. H. March, {\it Theoretical Solid State Physics}, (Dover, New York, 1985).


\end{thebibliography}
\end{document}